\documentstyle[prl,aps]{revtex}
\widetext
\newcommand{\beq}{\begin{equation}}
\newcommand{\eeq}{\end{equation}}
\newcommand{\bea}{\begin{eqnarray}}
\newcommand{\eea}{\end{eqnarray}}
\newcommand{\ba}{\begin{array}}
\newcommand{\ea}{\end{array}}

\def\to{\rightarrow}
\def\la{\lambda}
\def\l{\lambda}
\def\a{\alpha}
\def\matA{{\mbox{\bf A}}}

\begin{document}
\preprint{
NBI-HE-97-**\\
NSF-ITP-97-140\\
hep-th/9711096}
\draft
\title{Universal Massive Spectral Correlators and QCD$_{\bf 3}$}
\author{Poul H. Damgaard$^1$ and Shinsuke M. Nishigaki$^{2,*}$}
\address{
${}^{1}$The Niels Bohr Institute, Blegdamsvej 17, DK-2100 Copenhagen \O,
Denmark}
\address{
${}^2$Institute for Theoretical Physics, University of California,
Santa Barbara, CA 93106, USA}
\date{\today}
\maketitle
\begin{abstract} 
Based on random matrix theory in the unitary ensemble, we derive
the double-microscopic massive spectral correlators corresponding
to the Dirac operator of QCD$_3$ with an even number of fermions
$N_f$. We prove that these spectral correlators are universal, and
demonstrate that they satisfy exact massive spectral sum rules of QCD$_3$
in a phase where flavor symmetries are spontaneously broken 
according to $U(N_f)\!\to\! U(N_f/2)\!\times\!U(N_f/2)$.
\end{abstract}


%

It has long been suspected that QCD$_3$, and
even QED$_3$, with an {\em even} number of flavors 
$N_f\equiv 2\alpha$ may undergo spontaneous flavor symmetry breaking
according to the pattern $U(N_f)\!\to\! U(N_f/2)\!\times\!U(N_f/2)$
\cite{P}. This phenomenon can most easily be understood when one notices
that for an even number of flavors, the original two-spinors of
$(2\!+\!1)$-dimensional fermions may be grouped pairwise into half as
many four-spinors \cite{CJT}. The resulting formalism has an uncanny
resemblance to QCD$_4$ with $N_f/2$ flavors, and
{\em two} chiral symmetries, those associated with $\gamma_4$ and
$\gamma_5$ rotations \cite{P}. The global symmetry is, however, slightly
unusual: $U(N_f)$, as follows directly from the original formulation
in terms of two-spinors. The suggested flavor symmetry breaking can
be directly understood in terms of the pseudo-chiral symmetry described
above. Moreover, as has been remarked recently \cite{VZ}, the 
Coleman-Witten argument \cite{CW} applied to QCD$_3$ in the limit of 
many colors $N_c$ leads to precisely this prediction.

An order parameter for the above symmetry breaking pattern is the
absolute value of the chiral condensate, $\Sigma \!\equiv\! \sum_i\langle
\bar{\psi}_i\psi_i\rangle/N_f$. By an analogue of the Banks-Casher
relation \cite{BC}, this condensate is related to the spectral density
of the Dirac operator, evaluated at the origin, 
$\rho(0)=\Sigma/\pi$. A most
remarkable and testable prediction of Verbaarschot and Zahed \cite{VZ}
is that the massless QCD$_3$ spectral density $\rho(\la)$ near the origin at
$\la\!=\!0$, the microscopic spectral density
\beq
\rho_S(x) \equiv \lim_{V\!\to\!\infty} \frac{1}{V\Sigma}
\rho(\frac{x}{V\Sigma}) ~,
\eeq
may be computed {\em exactly} in a unitarily invariant
random matrix ensemble \cite{VZ}. 
Here $V$ denotes the three-volume, and the microscopic spectral density
is therefore to be considered as a finite-volume scaling function. In
this particular case the volume $V$ translates directly into the size
$N$ of random matrices in the unitary ensemble. 

The only required input for the above conjecture is the existence of a chiral 
condensate, and hence a non-vanishing $\rho(0)$. There is now substantial
evidence that the analogous statements for $(3\!+\!1)$-dimensional theories
\cite{V} are correct, ranging from agreement with exact massless spectral
sum rules of QCD and generalizations \cite{LS,SmV} to an explicit lattice
gauge theory computation of the microscopic spectral density \cite{BBMSVW}.
An essential ingredient in understanding how random matrix theory
can provide exact statements about full-fledged quantum field theories
is the proven universality, within random matrix theory, of the pertinent
microscopic spectral densities \cite{ADMN}.

Very recently the microscopic spectral densities of $SU(N_c\geq 3)$
gauge theories with $N_f$ massive flavors in $(3\!+\!1)$ dimensions have been
computed from random matrix theory \cite{DN} (see also \cite{WGW}).
Such an extension is essential for future
comparisons with lattice gauge theory beyond the quenched approximation.
Remarkably, also these double-microscopic spectral densities 
(called so because both eigenvalues and masses need to be rescaled with
volume $V$) are universal. Moreover, the double-microscopic massive
spectral densities satisfy exact massive spectral sum rules of QCD 
\cite{DN,D}.

In this letter we shall extend the computation of double-microscopic
massive spectral densities to the case of the random unitary 
invariant matrix ensemble,
which, in view of the work of Verbaarschot and Zahed \cite{VZ} can
provide exact information about the Dirac operator spectrum in QCD$_3$.
We shall prove that these double-microscopic spectral densities
(and spectral correlators) are universal
within the framework of random matrix theory, and show that they
satisfy exact massive spectral sum rules of QCD$_3$. In doing this, we
shall also provide an exact representation of the finite-volume
partition function for QCD$_3$ in the so-called mesoscopic range of
volumes \cite{V}.

Our starting point is the random matrix ensemble of
partition function
\beq
{\cal Z}~=~ \int\! dM \prod_{f=1}^{N_{f}}{\det}\left(M + im_f\right)~ 
{\rm e}^{-{N} {\rm tr}\, V(M^2)} ~.
\label{cZ}
\eeq
The integration is over hermitian $N\times N$ matrices $M$
with the associated Haar measure, and we parametrize
the potential in a general way by $
V(M^2) = \sum_{k\geq 1} (g_{k}/(2k)) M^{2k}$.
Masses are grouped pairwise with opposite signs \cite{VZ}:
$\{m_f\}=\{m_1,-m_1,m_2,-m_2,\ldots\}$.
Introducing the eigenvalues $\lambda_i$ of the hermitian
matrix $M$ we have, discarding an irrelevant overall factor,
\beq
{\cal Z}~=~\int_{-\infty}^{\infty}\! \prod_{i=1}^N \left(d\lambda_i
\prod_{f=1}^{N_f/2}(\lambda_i^2 + m_f^2)~
{\rm e}^{-NV(\lambda_i^2)}\right)\left|{\det}_{ij}
\lambda_j^{i-1}\right|^2 ~.
\eeq
We thus seek polynomials 
$P_n^{[\alpha]}(\la;m_1,\ldots,m_{\a})$ 
orthogonal with respect to the weight functions
\beq
w(\lambda) = \prod_{f=1}^{\a}(\lambda^2 + m_f^2)~
{\rm e}^{-NV(\la^2)} ~.
\label{weight}
\eeq
Since the weight functions are even in $\la$ due to the pairwise assignment
of masses, the polynomials split into even and odd sectors.
We treat these two sectors separately,
until we combine them to construct the kernel.
In ref.\ \cite{ADMN} it was proven that when all $m_f\!=\! 0$ the orthogonal 
polynomials have, for fixed $x \!=\! N\la$ and $t\!=\! 2n/N$, a universal 
limiting behavior. Normalized according
to $P_{2n}^{[\alpha]}(0) \!=\!{P_{2n+1}^{[\alpha]}}'(0) \!=\! 1$ the limit is 
\begin{mathletters}
\label{bessels}
\bea
{\cal P}_+^{[\a]}(t;x)&\equiv&
\lim_{N\to\infty} \left.P_{2n}^{[\alpha]}(\frac{x}{N})\right|_{n=Nt/2}
=
\Gamma(\alpha\!+\!\frac{1}{2})\frac{J_{\alpha-\frac{1}{2}}(u(t)x)}
{(u(t)x/2)^{\alpha-\frac{1}{2}}}~, 
\label{evenbessel}\\
{\cal P}_-^{[\a]}(t;x)& \equiv&
\lim_{N\to\infty} \left.{N}\,
P_{2n+1}^{[\alpha]}(\frac{x}{N})\right|_{n=Nt/2}
=
x\,\Gamma(\alpha\!+\!\frac{3}{2})\frac{J_{\alpha+\frac{1}{2}}(u(t)x)}
{(u(t)x/2)^{\alpha+\frac{1}{2}}} ~.
\label{oddbessel}
\eea
\end{mathletters}
Here $u(t)$ is determined by
\beq
u(t) ~=~ \int_0^t\frac{dt'}{2\sqrt{r(t')}} ~~,~~~~~~~~~
t ~=~ \sum_k \frac{g_{k}}{2}\left(2k \atop k \right) r(t)^k ~. 
\eeq
and $u(1)\!=\! \pi\rho(0)$
(where $\rho(0)$ is the large-$N$ spectral density at the origin).

We are now in a position to generalize this result
to the case of massive fermions, following
the method developed in ref.\ \cite{DN}.
We use the following lemma to 
construct the required polynomials:
\begin{quote}
{\bf Lemma 1 (Christoffel)} :
If $\{P_{n}(\lambda)\}_{n=0,1,\cdots}$ 
is a set of polynomials
orthogonal with respect to an {\em even} weight function $w(\lambda)$,
\beq
\tilde{P}_{n}(\lambda)
= \frac{P_{n}(\lambda)P_{n+2}(\lambda')-
        P_{n+2}(\lambda)P_{n}(\lambda')}{\lambda^2-{\lambda'}^2}
\label{Christ}
\eeq
are polynomials orthogonal with respect to
the weight $(\lambda^2-{\lambda'}^2)\,w(\lambda)$
\cite{SZE}.
\end{quote}
By replacing $\l'\to i m$,
we can use this procedure to incorporate the factor $\l^2+m^2$
due to a pair of fermions of masses $\pm m$
into the weight function.
By iterating this procedure,
we can construct polynomials orthogonal with respect to
the weight (\ref{weight}).
In the large-$n,\,N$ limit, the difference in $n$ in the numerator
of (\ref{Christ}) is replaced by the differential in $t$.
Then the next lemma allows us to express the polynomials
in a closed form:
\begin{quote}
{\bf Lemma 2} :
Let $P^{[\alpha]}(t;\lambda_0,\lambda_1,\cdots,\lambda_\alpha)$,
$\alpha=0,1,2,\cdots$, be
a set of functions generated by the iteration
\bea
&&P^{[\alpha+1]}(t;\lambda_0,\lambda_1,\cdots,\lambda_{\alpha+1})=\\
&&\frac{
P^{[\alpha]}(t;\lambda_0,\lambda_1,\cdots,\lambda_{\alpha})
P_t^{[\alpha]}(t;\lambda_{\alpha+1},\lambda_1\cdots,\lambda_{\alpha})-
P_t^{[\alpha]}(t;\lambda_0,\lambda_1,\cdots,\lambda_{\alpha})
P^{[\alpha]}(t;\lambda_{\alpha+1},\lambda_1\cdots,\lambda_{\alpha})}
{\lambda^2_0-\lambda^2_{\alpha+1}}~.\nonumber
\eea
Then they are given by
\beq
P^{[\alpha]}(t;\lambda_0,\lambda_1,\cdots,\lambda_{\alpha-1},\lambda_\alpha)=
c(t;\lambda_1,\cdots,\lambda_{\alpha})\,
\frac{\det_{i,j} P^{(i)}(t;\lambda_j)}
{
\prod_{i=1}^{\a} (\l_0^2-\lambda_i^2)
}
\label{lemma2}
\eeq
where 
$P^{(i)}(t;\lambda)=\frac{\partial^i}{\partial t^i}P^{[0]}(t;\lambda)$,
and
$c(t;\lambda_1,\cdots,\lambda_{\alpha})$ is a function in $t$ 
and in $\{\lambda_1,\cdots,\lambda_{\alpha}\}$.
\end{quote}
We refer the reader to ref.\ \cite{DN}
for the proof of these lemmata.
Now we replace 
$\lambda_i\to{x_i}/{N}$ and
$P^{[0]}(t;\l)$ by its microscopic limit (\ref{bessels}),
$P^{[0]}_{\pm}(t;{x}/{N}) \to 
u(t)^{\pm \frac12}x^{\frac12} J_{\mp\frac12} (u(t)x)$.
Here the upper and lower signs stand for polynomials in
the even and odd sectors, respectively.
Then we can prove by induction that 
its $t$-derivatives are expressed as
\beq
P^{(i)}_\pm (t;\frac{x}{N})
\to
\sum_{k=0}^i d_{i,k}(t)\,x^{k+\frac12}\,J_{k\mp\frac12}(u(t)x) ~.
\eeq
Once it is substituted inside the determinant
$\det\,P^{(i)}(\l_j)$, 
only the top term proportional to $
x^{i+\frac12}\,J_{i\mp\frac12}(u(t)x)
$
contributes. Thus the determinant in (\ref{lemma2}) is replaced by
\beq
d(t)\,
\det_{0\leq i,j\leq \alpha} 
x_j^{i+\frac12}\,J_{i\mp\frac12}(u(t)x_j)~.
\eeq
Performing the analytical continuation of $(\zeta_1,\cdots,\zeta_\a)$ 
to imaginary $(i\mu_1,\cdots,i\mu_\a)$, we thus obtain
the microscopic limit of the orthogonal polynomials:
\beq
{\cal P}^{[\a]}_{\pm}(t;x,\{\mu_f\})\equiv 
\lim_{N\to\infty}
P^{[\alpha]}_\pm(t; \frac{x}{N},\{\frac{\mu_f}{N}\})
=
c(t,\{\mu_f\})\,
\frac{
\left|
\begin{array}{cccc}
x^{\frac12}J_{\mp\frac12}(u(t)x) & 
x^{\frac32}J_{1\mp\frac12}(u(t)x) & 
\cdots & 
x^{\a+\frac12} J_{\a\mp\frac12}(u(t)x)\\
\mu_1^{\frac12}I_{\mp\frac12}(u(t)\mu_1) & 
-\mu_1^{\frac32} I_{1\mp\frac12}(u(t)\mu_1) & 
\cdots & 
(-)^\a \mu_1^{\a+\frac12} I_{\a\mp\frac12}(u(t)\mu_1)\\
\vdots & \vdots & \cdots & \vdots \\
\mu_\a^{\frac12}I_{\mp\frac12}(u(t)\mu_\a) & 
-\mu_\a^{\frac32} I_{1\mp\frac12}(u(t)\mu_\a) & 
\cdots & 
(-)^\a \mu_\a^{\a+\frac12} I_{\a\mp\frac12}(u(t)\mu_\a)
\end{array}
\right|
}{\prod_{f = 1}^{\alpha}(x^2+\mu^2_f)} ~.
\label{generalP}
\eeq
The microscopic kernel and spectral density
are constructed out of ${\cal P}^{[\alpha]}_\pm(t=1)$ 
as
\beq
K^{(\a)}(x,x';\{\mu_f\})=C(\{\mu_f\})
\sqrt{\prod_{f=1}^\a (x^2+\mu_f^2)({x'}^2+\mu_f^2)}
\frac{ {\cal P}^{[\alpha]}_+(1;x,\{\mu_f\})
       {\cal P}^{[\alpha]}_-(1;x',\{\mu_f\}) -
       {\cal P}^{[\alpha]}_-(1;x,\{\mu_f\})
       {\cal P}^{[\alpha]}_+(1;x',\{\mu_f\})
}{x-x'}~,
\label{kernel}
\eeq
\vspace{-8mm}
\bea
&&\rho^{(\a)}_S(x;\{\mu_f\})=\frac{1}{\pi\rho(0)}
K^{(\a)}\left(
\frac{x}{\pi\rho(0)},\frac{x}{\pi\rho(0)}
;\left\{\frac{\mu_f}{\pi\rho(0)}\right\} \right)
\nonumber\\
&&
=\frac{{\cal C}(\{\mu_f\})}{\prod_{\a=1}^f (x^2+\mu_f^2)}
\left(
\left|
\begin{array}{ccc}
x^{\frac12}J_{-\frac12}(x) & 
\cdots & x^{\a+\frac12} J_{\a-\frac12}(x)\\
\mu_1^{\frac12}I_{-\frac12}(\mu_1) & 
\cdots & (-)^\a \mu_1^{\a+\frac12} I_{\a-\frac12}(\mu_1)\\
\vdots & \cdots & \vdots \\
\mu_\a^{\frac12}I_{-\frac12}(\mu_\a) & 
\cdots & (-)^\a \mu_\a^{\a+\frac12} I_{\a-\frac12}(\mu_\a)
\end{array}
\right|
\left|
\begin{array}{ccc}
x^{\frac12}J_{-\frac12}(x) 
& 
\cdots &
x^{\a+\frac12} J_{\a-\frac12}(x)
\\
\mu_1^{\frac12}I_{\frac12}(\mu_1) & 
\cdots & 
(-)^\a \mu_1^{\a+\frac12} I_{\a+\frac12}(\mu_1)\\
\vdots & \cdots & \vdots \\
\mu_\a^{\frac12}I_{\frac12}(\mu_\a) & 
\cdots & 
(-)^\a \mu_\a^{\a+\frac12} I_{\a+\frac12}(\mu_\a)
\end{array}
\right| \right.
\label{rhoS}\\
&&-
\left.
\left|
\begin{array}{ccc}
x^{-\frac12}J_{-\frac12}(x) +
x^{\frac12}J_{-\frac32}(x) 
& 
\cdots &
x^{\a-\frac12} J_{\a-\frac12}(x)+
x^{\a+\frac12} J_{\a-\frac32}(x)
\\
\mu_1^{\frac12}I_{-\frac12}(\mu_1) & 
\cdots & (-)^\a \mu_1^{\a+\frac12} I_{\a-\frac12}(\mu_1)\\
\vdots & \cdots & \vdots \\
\mu_\a^{\frac12}I_{-\frac12}(\mu_\a) & 
\cdots & (-)^\a \mu_\a^{\a+\frac12} I_{\a-\frac12}(\mu_\a)
\end{array}
\right|
\left|
\begin{array}{ccc}
x^{\frac12}J_{\frac12}(x) & 
\cdots & 
x^{\a+\frac12} J_{\a+\frac12}(x)\\
\mu_1^{\frac12}I_{\frac12}(\mu_1) & 
\cdots & 
(-)^\a \mu_1^{\a+\frac12} I_{\a+\frac12}(\mu_1)\\
\vdots & \cdots & \vdots \\
\mu_\a^{\frac12}I_{\frac12}(\mu_\a) & 
\cdots & 
(-)^\a \mu_\a^{\a+\frac12} I_{\a+\frac12}(\mu_\a)
\end{array}
\right|
\right).
\nonumber
\eea
The constant ${\cal C}(\{\mu_f\})=
C(\{\mu_f\})/(\pi\rho(0))^{\a^2+1}$ 
is determined to be
\beq
{\cal C}(\{\mu_f\})^{-1}=2
\det_{1\leq i,j \leq \a}
\left((-)^i \mu^{i-\frac12}_j I_{i-\frac32}(\mu_j)\right)\,
\det_{1\leq i,j \leq \a}
\left((-)^i \mu^{i-\frac12}_j I_{i-\frac12}(\mu_j)\right)
\eeq
by requiring the matching between the 
$x\rightarrow\infty$ limit of the microscopic density
(normalized as in (\ref{rhoS}))
and the macroscopic density at $\l=0$:
\beq
\rho^{(\alpha)}_S (x\rightarrow\infty ; \{\mu_f\}) 
\to
\frac{1}{\pi}.
\eeq
For convenience we exhibit the first two examples of $\rho_S^{(\a)}$
(with degenerate masses $\mu$):
\begin{mathletters}
\bea
&&
\pi \rho_S^{(1)}(x;\mu)=
1+\frac{\mu}{x^2+\mu^2}\frac{\cos 2x - \cosh 2\mu}{\sinh 2\mu}~,
\\
&&
\pi\rho_S^{(2)}(x;\mu,\mu)=
1 - {\frac{\mu \left( 4\mu \left( {x^2} - {{\mu }^2} \right) 
         \left( 1 - \cos 2x \cosh 2\mu  \right)  + 
        2\left( \left( {x^2} + {{\mu }^2} \right) 
\left( \cos 2x - \cosh 2\mu  \right)  + 
           4x{{\mu }^2}\sin 2x \right) 
\sinh 2\mu  \right) }{{{\left( {x^2} + {{\mu }^2} \right) }^2}
\left( 4{{\mu }^2} - \sinh^2 2\mu  \right) }}~.
\eea
\end{mathletters}

It follows directly from the above construction and the universality
proof of ref. \cite{ADMN} that we have simultaneously proven that 
the orthogonal polynomials
(\ref{generalP}), the kernel (\ref{kernel}) (as well as all higher spectral 
correlators), and the microscopic spectral density itself (\ref{rhoS})
are {\em universal}, i.e. insensitive to the potential $V(M^2)$ in this limit.

By using the minor expansion of the determinants
and Hankel's asymptotic formula for the Bessel functions,
we can easily check that 
the microscopic kernels (\ref{kernel}) and densities (\ref{rhoS})
for arbitrary $\alpha$ satisfy a sequence of
decoupling relations for heavy fermions \cite{DN}:
\beq
\rho_S^{(\alpha)}(x;\mu_1,\ldots,\mu_{\alpha-1},\mu_{\alpha})
\stackrel{\mu_{\alpha} \to \infty}{\longrightarrow}
\rho_S^{(\alpha-1)}(x;\mu_1,\ldots,\mu_{\alpha-1})
\stackrel{\mu_{\alpha-1} \to \infty}{\longrightarrow}
\rho_S^{(\alpha-2)}(x;\mu_1,\ldots,\mu_{\alpha-2})
\to \cdots~.
\label{decouple}
\eeq
We similarly verify
that when all masses vanish, $\rho_S^{(\alpha)}(x;0,\cdots,0)$ 
agrees with the result obtained directly 
from the massless case \cite{VZ},
\beq
\rho^{(\alpha)}_S (x ; 0,\cdots,0) =
\frac{x}{4}\left(
 J_{\alpha+\frac12}(x)^2+J_{\alpha-\frac12}(x)^2
-J_{\alpha+\frac12}(x)J_{\alpha-\frac32}(x)
-J_{\alpha-\frac12}(x)J_{\alpha+\frac32}(x) \right) ~.
\label{masslessrho}
\eeq

It finally remains to compare these universal matrix model results with
exact massive spectral sum rules of QCD$_3$ in the phase of broken flavor
symmetry. In ref.\ \cite{VZ} it was argued that the relevant finite-volume
partition function for QCD$_3$ can be written
\beq
{\cal Z(M)} ~=~ \int dU 
\exp[N\Sigma\,{\rm tr}({\cal M}U\Gamma_5 U^{\dagger})] ~,
\label{qcd3}
\eeq
where the integration has been extended from the coset
$U(N_f)/U(N_f/2)\!\times\!U(N_f/2)$ to $SU(N_f)$. 
The mass matrix ${\cal M}$ 
takes the form ${\rm diag}(m_1,\cdots,m_{N_f/2},-m_1,\ldots,-m_{N_f/2})$.
The other matrix is $\Gamma_5\!=\! {\rm diag}(\openone,-\openone)$,
where $\openone$ is 
an $(N_f/2)\!\times\! (N_f/2)$ unit matrix. As could have been guessed by
comparison with the case of QCD$_4$ \cite{JSV}, the partition function
(\ref{qcd3}) is an example of the 
Harish-Chandra--Itzykson-Zuber integral
\cite{HCIZ}, 
now for hermitian matrices. The only slight complication
arises from the fact that $\Gamma_5$ has two sets of $N_f/2$-fold 
degenerate eigenvalues, which makes the standard expression for the
integral indeterminate. One can take care of this by regularizing the
$\Gamma_5$ matrix in any way that removes the degeneracy, performing
the integral, and subsequently taking the degenerate limit. We define
$\mu_i\!\equiv\!N\Sigma m_i$. Using the prescription above, 
the integral (\ref{qcd3}) can be performed explicitly, and one gets, 
up to an irrelevant normalization factor,
\beq
{\cal Z(M)} ~=~ \frac{\det\left(
\begin{array}{ll}
\matA(\{\mu_i\}) & \matA(\{-\mu_i\})\\
\matA(\{-\mu_i\})  & \matA(\{\mu_i\})
\end{array}\right)}{\Delta({\cal M})}
\eeq
where $\Delta({\cal M})$ is the Vandermonde determinant of the mass
matrix ${\cal M}$. The $(N_f/2)\!\times\!(N_f/2)$ matrix
$\matA(\{\mu_i\})$ is defined by $\matA_{ij}\!\equiv\! \mu_i^{j-1}
e^{\mu_i}$. An analogous procedure applies to the mass matrix 
if one insists on getting the result with some or all of the $N_f/2$ 
mass eigenvalues being equal. 
Massive spectral sum rules can now be derived by taking derivatives with
respect to one or more of the mass eigenvalues \cite{D}. For example, 
for 2 and 4 fermion species of degenerate (up to a sign, see the discussion
above) masses $m$, this gives for the simplest sum rules 
(summing over {\em positive} eigenvalues only)
\begin{mathletters}
\bea
N_f\!=\!2:& &\ \ 
\left\langle \sum_n \!~' \frac{1}{\la_n^2 + m^2} \right\rangle ~=~
\frac{\Sigma^2 N^2}{2\mu}\left(\coth 2\mu - \frac{1}{2\mu}\right) ~,\\
N_f\!=\!4:& &\ \ 
\left\langle \sum_n \!~' \frac{1}{\la_n^2 + m^2} \right\rangle ~=~
\frac{\Sigma^2 N^2}{2\mu^2} \frac{\sinh^2 2\mu  - {\mu} \sinh 2\mu
\cosh 2\mu - 2\mu^2}{4\mu^2 - \sinh^2 2\mu } ~.
\eea
\end{mathletters}
We note that in the limit 
$\mu\!\to\! 0$ these sum rules reduce correctly to those of the massless case
\cite{VZ}, where the right hand sides above are replaced by
$\Sigma^2 N^2 N_f/(2(N_f^2\!-\! 1))$.

We can now check these massive spectral sum rules by means of the identity
\beq
\frac{1}{N^2\Sigma^2}\left\langle \sum_n \!~' 
\frac{1}{\la_n^2 + m^2} \right\rangle ~=~
\int_0^{\infty} \! dx ~\frac{\rho_S^{({N_f}/{2})}
(x;\mu,\ldots,\mu)}{x^2+\mu^2} ~.
\eeq
and the general expression (\ref{rhoS}). The integrals are elementary, 
and we find that the massive spectral sum rules are exactly satisfied.

\end{document}